\documentclass[10pt,letterpaper]{article}
\usepackage[top=0.85in,left=2.75in,footskip=0.75in]{geometry}
\usepackage{amsmath,amssymb}
\usepackage{changepage}
\usepackage[utf8x]{inputenc}
\usepackage{textcomp,marvosym}
\usepackage{cite}
\usepackage{nameref,hyperref}
\usepackage{microtype}
\DisableLigatures[f]{encoding = *, family = * }
\usepackage[table]{xcolor}
\usepackage{array}
\newcolumntype{+}{!{\vrule width 2pt}}
\newlength\savedwidth

\raggedright
\usepackage[aboveskip=1pt,labelfont=bf,labelsep=period,justification=raggedright,singlelinecheck=off]{caption}

\makeatletter
\renewcommand{\@biblabel}[1]{\quad#1.}
\makeatother
\usepackage{lastpage,fancyhdr,graphicx}
\usepackage{epstopdf}
\fancyheadoffset[L]{2.25in}
\fancyfootoffset[L]{2.25in}
\usepackage[T1]{fontenc}
\usepackage{chemfig}
\usepackage{chemformula}
\usepackage{amsmath,bm}
\usepackage{braket}
\usepackage{xcolor}
\begin{document}
\vspace*{0.2in}
\begin{flushleft}
{\Large
\textbf\newline{Radical pairs may explain reactive oxygen species-mediated effects of hypomagnetic field on neurogenesis} 
}
\newline
\\
Rishabh\textsuperscript{1,2,3*},
Hadi Zadeh-Haghighi\textsuperscript{1,2,3},
Dennis Salahub\textsuperscript{1,2,4,5},
Christoph Simon\textsuperscript{1,2,3*}
\\
\bigskip
\textbf{1} Department of Physics and Astronomy, University of Calgary, Calgary, AB, Canada
\\
\textbf{2} Institute for Quantum Science and Technology, University of Calgary, Calgary, AB, Canada
\\
\textbf{3} Hotchkiss Brain Institute, University of Calgary, Calgary, AB, Canada
\\
\textbf{4} Department of Chemistry, University of Calgary, Calgary, AB, Canada.
\\
\textbf{5} Centre for Molecular Simulation, University of Calgary, Calgary, AB, Canada.
\bigskip
* rishabh1@ucalgary.ca (R); csimo@ucalgary.ca (CS)

\end{flushleft}

\section*{Abstract}
Exposures to a hypomagnetic field can affect biological processes. Recently, it has been observed that hypomagnetic field exposure can adversely affect adult hippocampal neurogenesis and hippocampus-dependent cognition in mice. In the same study, the role of reactive oxygen species (ROS) in hypomagnetic field effects has been demonstrated. However, the mechanistic reasons behind this effect are not clear. This study proposes a radical pair mechanism based on a flavin-superoxide radical pair to explain the modulation of ROS production and the attenuation of adult hippocampal neurogenesis in a hypomagnetic field. The results of our calculations favor a singlet-born radical pair over a triplet-born radical pair. Our model predicts hypomagnetic field effects on the triplet/singlet yield of comparable strength as the effects observed in experimental studies on adult hippocampal neurogenesis. Our predictions are also in qualitative agreement with experimental results on superoxide concentration and other observed ROS effects. We also predict the effects of applied magnetic fields and oxygen isotopic substitution on adult hippocampal neurogenesis. Our findings strengthen the idea that nature might harness quantum resources in the context of the brain.

\section*{Author summary}
Exposure to magnetic fields influences many neurobiological processes. The formation of new neurons (neurogenesis) in the hippocampal region of the adult brain plays a crucial role in learning and memory. It can be adversely affected by shielding the earth's magnetic field, and this effect is intimately related to ROS concentration.  In this study, we have developed a quantum mechanical model to explain this magnetic field dependence of adult hippocampal neurogenesis. Our model is also consistent with the observed ROS effects.


\section*{Introduction}
Despite its great successes, many essential questions are still unanswered in neuroscience~\cite{adolphs2015unsolved}, including the underlying principles of memory and learning, the workings of general anesthesia, computational properties of the brain and, most fundamentally, the generation of subjective experience~\cite{koch2016neural}. Therefore, it is crucial to explore whether the brain also uses quantum resources apart from the known classical ones. In recent decades, researchers have tried to understand some of the enigmatic biological processes such as animal magnetoreception, photosynthetic energy capture, olfaction, and consciousness using inherently quantum concepts such as superposition and entanglement~\cite{ball2011physics,lambert2013quantum,mohseni2014quantum,mcfadden2016life,kim2021quantum,adams2020quantum}. These results encourage the view that some quantum processes may be going on in the brain and may help answer some of the previously unanswered questions in neuroscience.
\par
Exposure to magnetic fields influences many neurobiological processes in various animals, from insects to human beings, for example, repetitive transcranial magnetic stimulation at low-intensity (LI-rTMS) induces axon outgrowth and synaptogenesis in mice~\cite{dufor2019neural}, drosophila’s circadian clock can be perturbed by magnetic fields~\cite{yoshii2009cryptochrome}, and extremely low-frequency magnetic fields have been shown to induce human neuronal differentiation through NMDA receptor activation~\cite{ozgun2019extremely}. Apart from all these magnetic field effects, several isotopic effects have also been observed in the brain, such as in the anesthetic effects of Xenon~\cite{li2018nuclear} and the behavioral effects of Lithium~\cite{ettenberg2020differential}. As shown in Refs~\cite{smith2021radical,zadeh2021entangled}, these isotopic effects may be interpreted as being due to the magnetic field of the nuclear spin. 

Researchers have also shown that the cellular production of ROS is magnetically sensitive~\cite{usselman2014spin,usselman2016quantum,pooam2020hek293,van2019weak}. ROS are biologically very vital chemical species. Studies have shown the importance of ROS in cellular signaling~\cite{zhang2016ros}. ROS play an essential role in various cellular activities such as cell proliferation, differentiation, migration, survival, and autophagy~\cite{terzi2020role}. Oxidative stress, an imbalance between production and accumulation of ROS, has been implicated in several psychological disorders~\cite{salim2014oxidative}. Studies have also shown that ROS play a role in neurogenesis and neuronal differentiation~\cite{vieira2011modulation,zhang2021long}.
\par
Adult hippocampal neurogenesis plays a crucial role in learning and memory and is sensitive to various external stimuli. Recently, Zhang et al.~\cite{zhang2021long} have shown that mice exposed to a hypomagnetic field (HMF) by shielding the geomagnetic field (GMF, present-day intensity value $25-65$ $\mu T$) show decreased neurogenesis in the hippocampal region, which results in impaired cognition. An HMF is a static field with an intensity lower than $5$ $\mu T$. An organism can be exposed to HMF during long-term deep-space flights and in artificial environments on earth, such as magnetically shielded rooms.
\par
 Zhang et al.~\cite{zhang2021long} show that long-term exposure to HMF impaired neurogenesis through decreasing adult neuronal stem cell proliferation, altering cell lineages in critical development stages of neurogenesis, impeding dendritic development of newborn neurons in the adult hippocampus, and resulting in impaired cognition. Using transcriptome analysis in combination with endogenous ROS \emph{in situ} labeling via hydroethidine, they revealed reduced levels of ROS in HMF-exposed mice. Zhang et al.~\cite{zhang2021long} have also shown that pharmacological inhibition of ROS removal via diethyldithiocarbamate (DDC) rescues defective adult hippocampal neurogenesis in HMF-exposed mice and that the inhibition of ROS production via apocynin (APO) blocks the rescue effect of a return to GMF on defective adult hippocampal neurogenesis in HMF exposed mice. Based on these observations, they concluded that reduced levels of ROS are responsible for HMF effects on adult hippocampal neurogenesis on a mechanistic level.
 \par
 In recent years, the radical pair mechanism~\cite{closs1969mechanism,steiner1989magnetic,hayashi2004introduction} (RPM) has been proposed as a quantum mechanical explanation to several magnetically sensitive biological phenomena~\cite{hore2016radical,usselman2016quantum,smith2021radical,zadeh2021entangled,zadeh2021radical,zadeh2021radicalB}. The canonical example of such an explanation is that of magnetoreception in migratory birds~\cite{hore2016radical}. This model relies on the coherent spin dynamics of pairs of radicals involving the cryptochrome (CRY) protein~\cite{ritz2000model}. Later, similar mechanisms were proposed for other migratory animals~\cite{mouritsen2018long}.  It is natural to ask whether a similar explanation could also be given to some of the magnetically sensitive phenomena happening in the brain. Recently, RPM-based mechanisms, not all involving CRY, have been proposed to explain the isotope effects on Xenon-induced anesthesia~\cite{smith2021radical} and Lithium effects on hyperactivity~\cite{zadeh2021entangled}, as well as the magnetic field effects on Drosophila’s circadian clock~\cite{zadeh2021radical} and microtubule reorganization~\cite{zadeh2021radicalB}.
 \par
 In the RPM, a pair of radicals (molecules that contain unpaired electrons) are created simultaneously, such that the spins on the two electrons, one on each radical, evolve coherently~\cite{brocklehurst1976spin,bagryansky2007quantum,fay2019quantum}. Radical pairs (RPs) are usually created in either singlet or triplet states. Nuclear spins in the neighborhood of radicals and an external magnetic field interact with the RP via hyperfine (HF) interactions and the Zeeman interaction, respectively. Due to HF interactions, neither singlet nor triplet states are, in general, eigenstates of the spin Hamiltonian. Therefore, the RP is initially in a non-stationary superposition which evolves coherently and causes singlet-triplet interconversion at frequencies determined by the HF and Zeeman interactions. Therefore, altering the external magnetic field or substituting an isotope with a different spin can change the extent and timing of the singlet-triplet interconversion, resulting in altered yields of products formed spin-selectively from the singlet and triplet states~\cite{brocklehurst1976spin,timmel1996oscillating,timmel1998effects,bagryansky2007quantum,fay2019quantum}.
\par
The conclusion of Ref~\cite{zhang2021long} that HMF effects manifest via a change in ROS concentration implies that ROS production is magnetically sensitive. Researchers have already observed oscillating magnetic field effects on ROS production. Usselman et al.~\cite{usselman2014spin,usselman2016quantum} have observed that oscillating magnetic fields at Zeeman resonance alter relative yields of cellular superoxide (\ch{O2^{.-}}) and hydrogen peroxide (\ch{H2O2}) ROS products in human umbilical cells.
 \par
 ROS's two primary cellular sources are the mitochondrial electron transport chain, and an enzyme family termed NADPH oxidase (Nox)\cite{wallace2010mitochondrial,terzi2020role,zhao2019mitochondrial}. Flavin-containing enzymes are known to play a role in mitochondrial ROS production~\cite{fedoseeva2017role}. Mitochondria-based ROS production could involve magnetically sensitive flavin and superoxide-based RP. Nox enzymes are present in several subcellular locations, including mitochondria in various cell types, including in the hippocampus\cite{terzi2020role}. The catalytic core of Nox enzymes is composed of six $\alpha$-helical transmembrane domains. The sixth transmembrane domain is linked to an intracellular FAD-binding domain via a segment, and the FAD-binding domain is linked to an NADPH binding domain at the C-terminus.  Nox enzymes transport electrons from NADPH, through FAD, across the plasma membrane to molecular oxygen (\ch{O2}) to generate superoxide~\cite{terzi2020role}. This electron transfer as well as the magnetic field dependence of ROS production naturally suggests an underlying flavin and superoxide-based RPM for ROS production and hence also as a basis of HMF effects on adult hippocampal neurogenesis.
 \par 
 CRY can be an alternative source of magnetically sensitive flavin and superoxide-based RP. The role of CRYs in many brain processes is well established~\cite{dufor2019neural,emery2000drosophila}. As mentioned earlier, the RPM model for avian magnetoreception relies on the coherent spin dynamics of pairs of radicals involving CRY protein, which contains the flavin adenine dinucleotide (FAD)~\cite{ritz2000model,hore2016radical}. The canonical RP thought to be produced from CRY is in the form of flavosemiquinone radical (\ch{FAD^{.-}}) and terminal tryptophan radical (\ch{TrpH^{.+}})~\cite{giovani2003light,hong2020electron,hore2016radical}. However, considerable evidence suggests an alternative RP with the superoxide radical, \ch{O2^{.-}} as a partner for the flavin radical, with \ch{FADH^{.}} and \ch{O2^{.-}} acting as the donor and the acceptor, respectively~\cite{hogben2009possible,muller2011light,lee2014alternative}. 
 \par
 Usselman et al.~\cite{usselman2016quantum} have already proposed a flavin and superoxide-based RPM to explain the effects of oscillating magnetic fields at Zeeman resonance on ROS production. Nox enzyme, mitochondria, and CRY can all be the sources of this flavin-superoxide RP. Similar RP has also been in the RPM-based explanation for isotopic dependence of behavioral effects of Lithium~\cite{zadeh2021entangled} and the magnetic field effects on the circadian clock~\cite{zadeh2021radical}. In these studies both singlet-born~\cite{zadeh2021entangled,zadeh2021radical} and triplet-born~\cite{hogben2009possible,muller2011light,lee2014alternative,usselman2016quantum} RPs have been considered.
\par

Zhang et al.~\cite{zhang2021long} suggest that the RPM cannot provide a viable explanation for their observations, but 
here we show that the RPM could be the underlying mechanism behind the HMF effects on adult hippocampal neurogenesis. We have proposed a RPM model based on flavin-superoxide RP to explain the modulation of ROS production and the resulting attenuation of adult hippocampal neurogenesis in the hypomagnetic field. The theoretical predictions of our model are compared with the experimental results of Zhang et al.~\cite{zhang2021long}, and we obtained effects of comparable size. In this study, we have considered both singlet-born and triplet-born RP and our calculations show that singlet-born RP is in agreement with the experimental results, whereas  triplet-born RP is in disagreement with the experiments.

\section*{Results}
\subsection*{Hypomagnetic field effects on adult hippocampal neurogenesis and Radical Pair Mechanism}
\subsubsection*{Quantifying HMF effects on adult hippocampal neurogenesis}
Zhang et al.~\cite{zhang2021long}, have produced qualitative and quantitative results to show the adverse effects of HMF on adult hippocampal neurogenesis. They have found that long-term exposure to HMF impaired neurogenesis through decreasing adult neural stem cells (aNSCs) proliferation, altering cell lineages in critical development stages of neurogenesis, and impeding dendritic development of newborn neurons in the adult hippocampus. In this study, we will use the reduction in aNSCs proliferation as the quantitative measure of the HMF effects on adult hippocampal neurogenesis. Zhang et al.~\cite{zhang2021long}, tested the effect of exposure to HMF on aNSC proliferation in the dentate gyrus (DG) of adult mice. The mice were injected with bromodeoxyuridine (BrdU) and sacrificed 2 hr later to examine aNSC proliferation during exposure to GMF or HMF. 
\par
Compared with GMF groups, there was a decrease in the numbers of BrdU+ labeled proliferating cells in the DG of HMF-exposed mice at 2 weeks of HMF exposure and beyond. For mice exposed to HMF ($0.29\pm0.01$ $\mu T$~\cite{zhang2021long}) for 8 weeks, the average number of BrdU+ cells was 1363.51, with a standard deviation of 131.34. Whereas, for the control (mice exposed to GMF, $55.26\pm0.05$ $\mu T$~\cite{zhang2021long}), the average number of BrdU+ cells after the same time was found to be 1826.20, with a standard deviation of 174.55. A decrease of 25.37$\%$ is observed due to 8 weeks of exposure to HMF (See Table~\ref{tab:avg. BrdU+ cells}).

\begin{table}[!ht]
\centering
\caption{\label{tab:avg. BrdU+ cells} The numbers of BrdU+ cells in mice after a 8 week exposure to HMF and GMF.} 
\begin{tabular}{|l|c|c|}
\hline
 & Average number of BrdU+ cells (8 week) & Stdev. \\
\hline
GMF & 1826.20 & 174.55
  \\
\hline
HMF & 1363.51 & 131.34  \\
\hline
Ratio & 1.34 & 0.18  \\
\hline
\end{tabular}

\end{table}

The ratio of numbers of BrdU+ cells at GMF to that of HMF ($1.34 \pm 0.18$) will serve as a measure of the strength of HMF effects. We will compare HMF effects predicted by our model against this experimental measure. Zhang et al.~\cite{zhang2021long} also observed that HMF exposure causes a decrease in superoxide concentration, and we will look for qualitative agreement between our model and this result. Zhang et al.~\cite{zhang2021long} have also shown that pharmacological inhibition of ROS removal rescues defective adult hippocampal neurogenesis in HMF-exposed mice and that the inhibition of ROS production blocks the rescue effect of a return to GMF on defective adult hippocampal neurogenesis in HMF exposed mice. We will also look for agreement between our model and these observed effects of inhibition of ROS production and removal.

\par
As our theoretical model will predict the fractional superoxide yield and Zhang et al.~\cite{zhang2021long} have quantified the superoxide levels by hydroethidine fluorescence measurements, the quantitative comparison of superoxide's theoretical and experimental values seems a natural check to our model. However, the large spread and non-normal distribution of data make the superoxide measurement unsuitable for such quantitative analysis. Therefore, we have only demanded a qualitative agreement between the theory and experiment as far as superoxide is concerned, i.e., the concentration of superoxide should decrease due to HMF exposure.
  
\subsubsection*{RPM model}
\begin{figure}[!ht]
\centering
\includegraphics[]{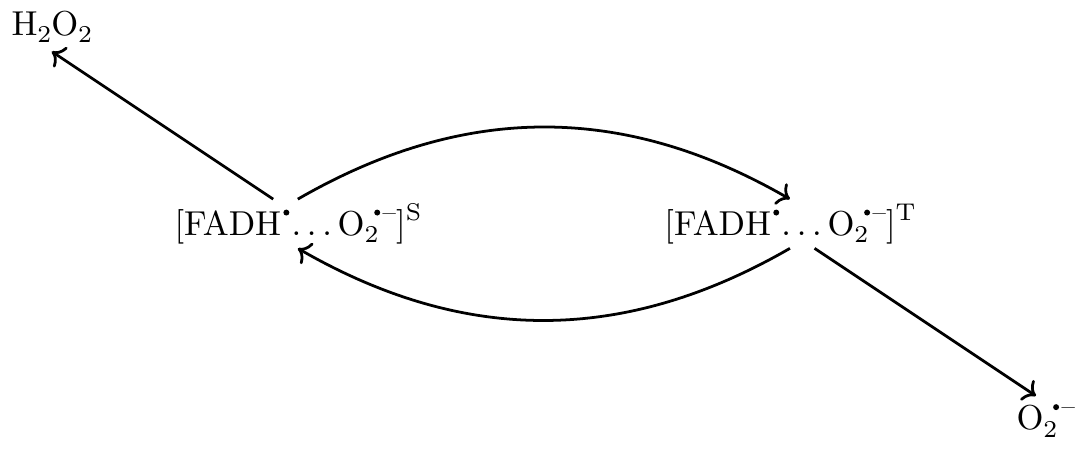}
\caption{{\bf Schematic of radical pair mechanism for \ch{[FADH^{.}... O2^{.-}]} radical pair.}}
\label{fig:RPM}
\end{figure}

We here propose an RP system of \ch{FADH^{.}} and \ch{O2^{.-}} to reproduce the HMF effects on adult hippocampal neurogenesis based on changes in the singlet-triplet yields at HMF as compared to GMF. The correlated spins of RP are taken to be in the \ch{[FADH^{.}... O2^{.-}]} form. The singlet-product is \ch{H2O2} and the triplet-product is \ch{O2^{.-}}~\cite{usselman2016quantum} (See Fig~\ref{fig:RPM}).
\par

\par
\begin{figure}[!ht]
\centering
\includegraphics[]{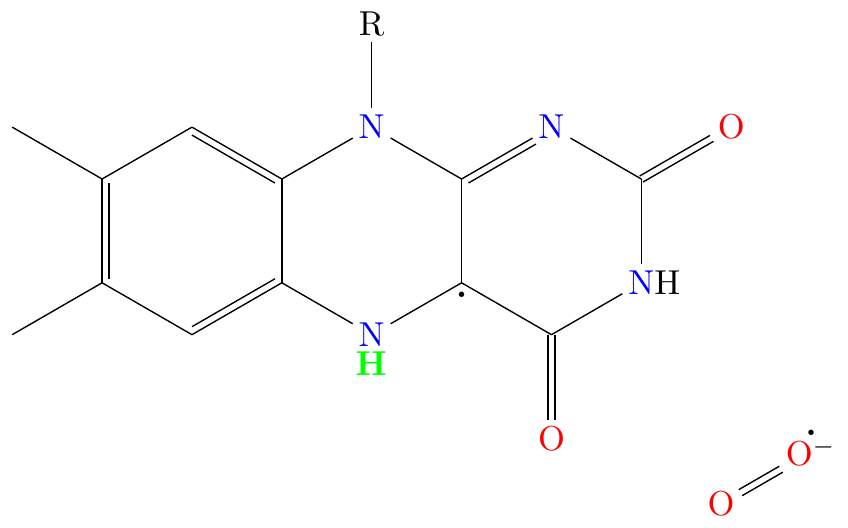}
\caption{{\bf \ch{[FADH^{.}... O2^{.-}]} RP}. H5 nucleus, which has the largest magnitude for isotropic HFCC among all the nuclei in \ch{FADH^{.}} is highlighted in green.}

\label{fig:RP}
\end{figure}

We consider a simple spin Hamiltonian for our RP by only including Zeeman and HF interactions~\cite{hore2016radical,efimova2008role}. Given the possible randomized orientation of the molecules involved, we only consider the isotropic Fermi contact contributions for the HF interactions. In our calculations, we assume that the unpaired electron on \ch{FADH^{.}} couples only with the H5 nucleus (See Fig~\ref{fig:RP}), which has the most prominent isotropic HF coupling constant (HFCC) among all the nuclei in \ch{FADH^{.}}~\cite{lee2014alternative}. The other unpaired electron, on \ch{O2^{.-}} (containing two \ch{^{16}O} nuclei), has no HF interactions. The Hamiltonian for our RP system reads as follows:
\begin{equation}\label{eq:ham}
\centering
\hat{H} = \omega \hat{S}_{{A}_{z}}+ a_{1}\mathbf{\hat{S}}_{A}.\mathbf{\hat{I}_{1}} + \omega \hat{S}_{{B}_{z}}, 	
\end{equation}
where $\mathbf{\hat{S}}_{A}$ and $\mathbf{\hat{S}}_{B}$ are the spin operators of radical electron $A$ and $B$, respectively, $\mathbf{\hat{I}_{1}}$ is the nuclear spin operator of the H5 of \ch{FADH^{.}}, $a_{1}$ is the HFCC between the H5 of \ch{FADH^{.}} and the radical electron A ($a = -802.9$ $\mu T$~\cite{lee2014alternative}), and $\omega$ is the Larmor precession frequency of the electrons due to the Zeeman effect.
\par
For simplicity, both the triplet and singlet reaction rates are identical and equal to $k$. The coherence lifetime of RP is taken to be $1/r$. 
\par
As mentioned earlier \ch{[FADH^{.}... O2^{.-}]} RP can be taken to start in either singlet~\cite{zadeh2021entangled,zadeh2021radical} or triplet states~\cite{hogben2009possible,muller2011light,lee2014alternative,usselman2016quantum} . In this study, we have considered both singlet and triplet-born RPs.

\subsubsection*{Singlet-born radical pair}
Let us first consider the initial state of the RP to be a singlet:
\begin{equation}\label{eq:singlet}
	\ket{S}\bra{S}\otimes \frac{1}{M}I_{M},
	\end{equation}
where $\ket{S}$ represents the singlet state of two electrons in RP, $M$ is the total number of nuclear spin configurations, and $\hat{I}_{M}$ is a $M$-dimensional identity matrix representing the completely mixed state of the nuclei.  
\par
The fractional triplet yield produced by the RPM can be obtained by tracking the spin state of the RP throughout the reaction~\cite{timmel1998effects,hore2019upper}. The ultimate fractional triplet yield ($\Phi_{T}$) for time periods much greater than the RP lifetime is:
\begin{equation}\label{eq:singlet triplet yield}
	{\frac{3}{4} + \frac{k}{4(k+r)} - \frac{1}{M}\sum_{m,n=1}^{4M}\frac{|{\bra{m}\hat{P}_{S}\ket{n}}|^{2}k(k+r)}{(k+r)^2+(\omega_{m}-\omega_{n})^{2}}},
	\end{equation}
where $\hat{P}^S$ is the singlet projection operator, $\ket{m}$and $\ket{n}$ are eigenstates of $\hat{H}$ (Eq.~\ref{eq:ham}) with corresponding eigenenergies of $\omega_m$ and $\omega_n$, respectively, $k$ is the RP lifetime rate, and $r$ is the RP spin-coherence lifetime rate.
\par
The fractional singlet yield ($\Phi_{S}$) can be calculated as:

\begin{equation}\label{eq:singlet singlet yield}
\Phi_{S} =	1-\Phi_{T}.
	\end{equation}
\par
\begin{figure}[!ht]
\centering
\includegraphics[scale=1]{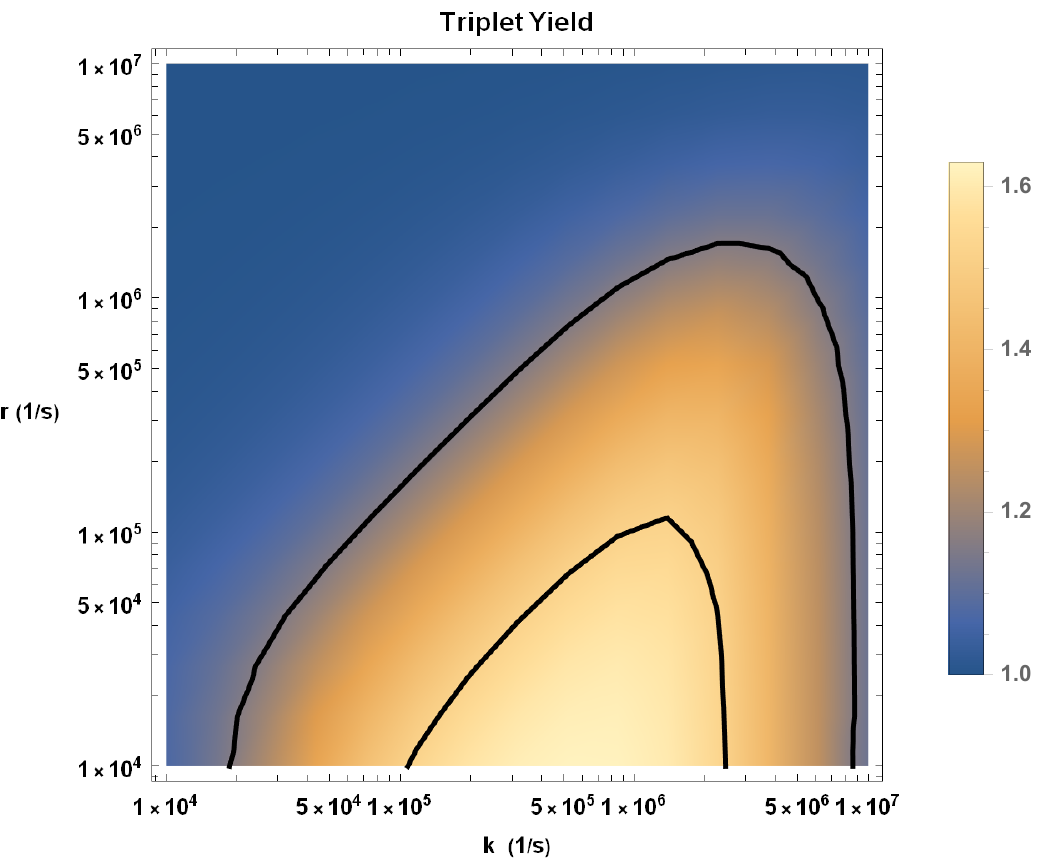}
\caption{{\bf Triplet yield ratio (GMF to HMF) for singlet-born RP in the $k-r$ plane.} The region between the solid black lines ($1.16$ and $1.52$) is in agreement with the experimental range for the ratio of the numbers of BrdU+ cells after an 8 week exposure to GMF and HMF($1.34\pm0.18$)~\cite{zhang2021long}. The value of HFCC $a_{1}$ is taken to be $a_{1} = -802.9$ $\mu T$~\cite{lee2014alternative}.}
\label{fig:triplet 1}
\end{figure}

We explore the effects of HMF on singlet (triplet) yield by calculating the singlet (triplet) yield ratio at GMF to HMF. We have plotted the triplet yield ratio on the $k$ and $r$ plane in Fig.~\ref{fig:triplet 1}. A triplet yield ratio greater than 1 implies that the concentration of superoxide decreases due to HMF exposure, which is in qualitative agreement with the experimental results. It also implies that the concentration of \ch{H2O2} increases due to HMF exposure, which is something not measured by Zhang et al.~\cite{zhang2021long}. For the quantitative comparison of the strength of effects of HMF exposure between the experimental measurements and our RPM model, the triplet and singlet yield ratios can be compared with the ratio of the numbers of BrdU+ cells after an 8 week exposure to GMF and HMF. The region between solid black lines in Fig.~\ref{fig:triplet 1} is in agreement with the experimental range of this measure, $1.34\pm0.18$. 
\par
As seen in Fig.~\ref{fig:triplet 1}, for values of $r$ and $k$ that are consistent with what is typically considered in the context of the RPM, the experimental effects of HMF agree with theoretical predictions.
\par
\begin{figure}[!ht]
\centering
\includegraphics[scale=1]{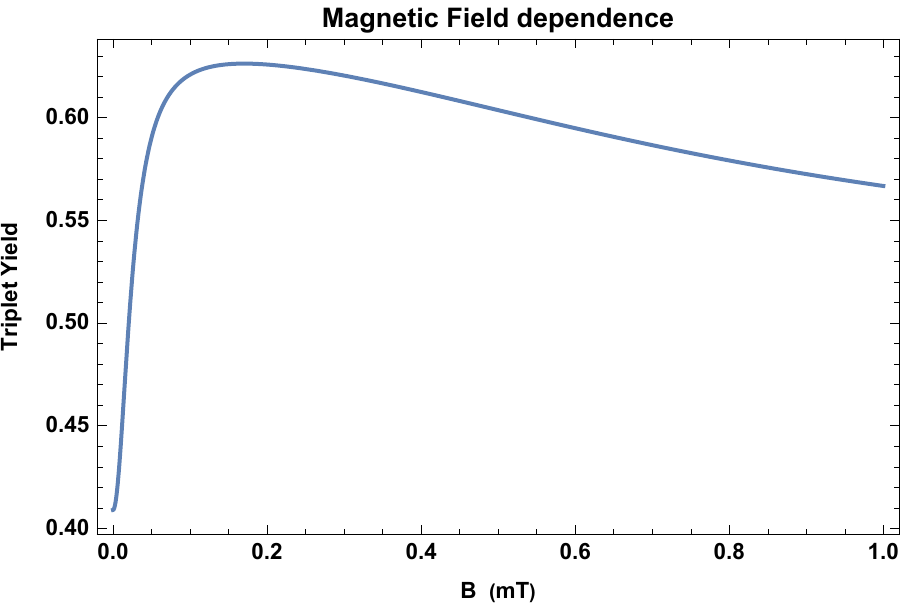}
\caption{{\bf Magnetic field dependence of $\Phi_{T}$.} The dependence of the triplet yield of the RPM for singlet-born radical pair on external magnetic field B for $a_{1} = -802.9$ $\mu T$~\cite{lee2014alternative}, reaction rate $k = 2\times 10^{6}$ $s^{-1}$, and relaxation rate $r = 2\times 10^{5}$ $s^{-1}$.}
\label{fig:mag}
\end{figure}

In a RPM the triplet (singlet) yield ratio can be altered by applying external magnetic fields. The dependence of the triplet yield of the RPM on external magnetic field for a singlet-born radical pair is shown in Fig.~\ref{fig:mag}.

\subsubsection*{Triplet-born radical pair}
For comparison, we now consider the initial state of the RP to be a triplet:
\begin{equation}\label{eq:columreal}
	\frac{1}{3}\Big{\{}\ket{T_{0}}\bra{T_{0}}+\ket{T_{+1}}\bra{T_{+1}}+\ket{T_{-1}}\bra{T_{-1}}\Big{\}}\otimes \frac{1}{M} I_{M},
	\end{equation}
where $\ket{T_{0}}$ and $\ket{T_{\pm1}}$ represent the triplet states of two electrons in RP with $m_{S}=0$ and $m_{S}=\pm1$ respectively.   
\\
The ultimate relative triplet yield ($\Phi_{T}$) for time periods much greater than the RP lifetime is:
\begin{equation}\label{eq:triplet triplet yield}
	{\frac{3}{4} - \frac{k}{12(k+r)} + \frac{1}{3M}\sum_{m,n=1}^{4M}\frac{|{\bra{m}\hat{P}_{S}\ket{n}}|^{2}k(k+r)}{(k+r)^2+(\omega_{m}-\omega_{n})^{2}}},
	\end{equation}
where the symbols have the above-stated meanings.
\begin{figure}[!ht]
\centering
\includegraphics[scale=1]{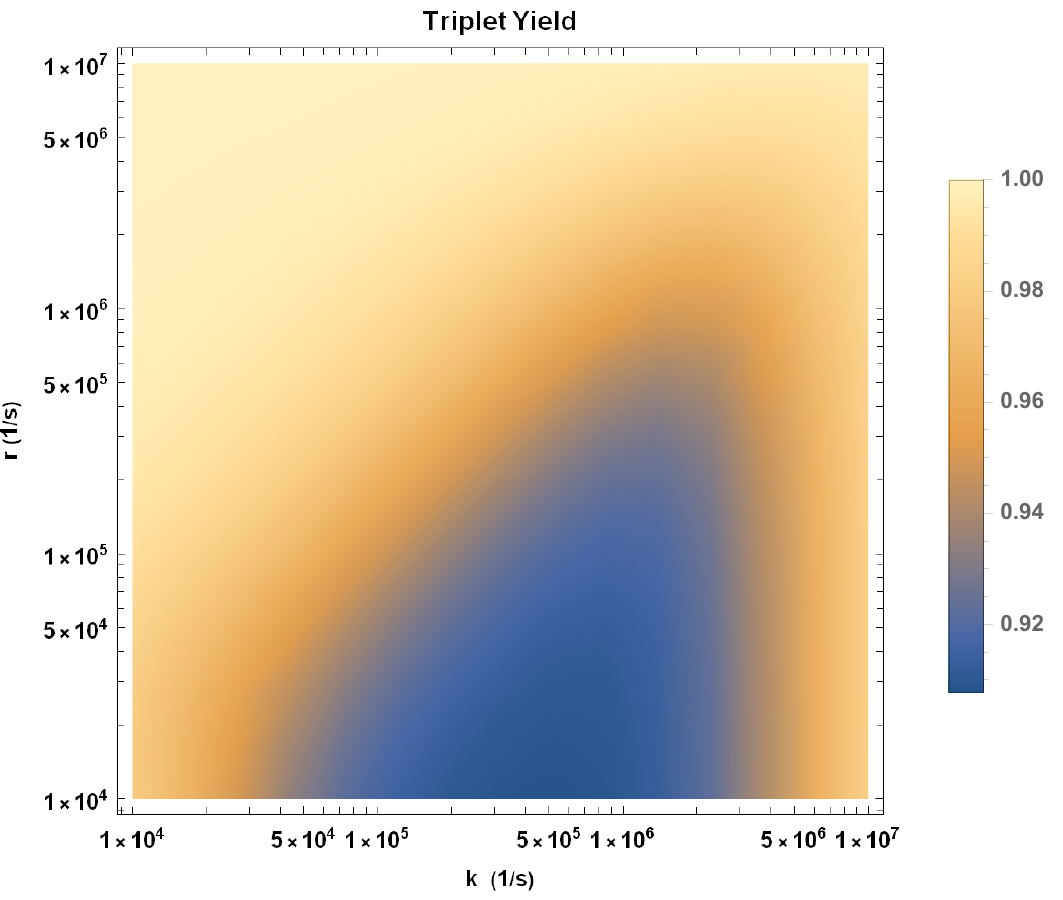}
\caption{{\bf Triplet yield ratio (GMF to HMF) for triplet-born RP in the $k-r$ plane.} The value of HFCC $a_{1}$ is taken to be $a_{1} = -802.9$ $\mu T$~\cite{lee2014alternative}.}
\label{fig:triplet 2}
\end{figure}
\par
As for the singlet-born case, we explore the effects of HMF on singlet (triplet) yield by calculating the singlet (triplet) yield ratio at GMF to HMF. We have plotted the triplet yield ratio on the $k$ and $r$ plane in Fig.~\ref{fig:triplet 2}. A triplet yield ratio less than 1 implies that the concentration of \ch{H2O2} decreases due to HMF exposure, which, as mentioned above, is not measured by Zhang et al.~\cite{zhang2021long}. It also implies that the concentration of superoxide increases due to HMF exposure, which disagrees with the experimental results. Therefore, a triplet-born RP is ruled out as an explanation for HMF effects on adult hippocampal neurogenesis.
\par
In summary, our calculations show that for a singlet-born \ch{[FADH^{.}... O2^{.-}]} RP, a decline is observed in the \ch{O2^{.-}} production upon HMF exposure and HMF effect of similar strength (defined earlier) as observed by Zhang et al.~\cite{zhang2021long} can be achieved for typical values of $k$ and $r$.

\subsubsection*{Isotopic effects of Oxygen}
Isotope effects can be an indication of the RPM\cite{smith2021radical,zadeh2021entangled}. Naturally occurring oxygen is almost exclusively found in the form of the \ch{^{16}O} isotope, which has a zero spin. If one of the oxygen atoms in superoxide radical is replaced with \ch{^{17}O}, which has a spin , $I_{2} = 5/2$, an additional HF term must be added to the RP Hamiltonian. The new Hamiltonian for our RP system reads as follows:
\begin{equation}\label{eq:ham2}
\centering
\hat{H} = \omega \hat{S}_{{A}_{z}}+ a_{1}\mathbf{\hat{S}}_{A}.\mathbf{\hat{I}_{1}} + \omega \hat{S}_{{B}_{z}} + a_{2}\mathbf{\hat{S}}_{B}.\mathbf{\hat{I}_{2}}, 	
\end{equation}
$\mathbf{\hat{I}_{2}}$ is the nuclear spin operator of the \ch{^{17}O}, $a_{2}$ is the HFCC between the \ch{^{17}O} and the radical electron B. We used density functional theory (DFT) to estimate the value of $a_{2}$ to be $1886.8$ $\mu T$ (see the methods section). The fractional triplet and singlet yields for singlet-born RP can be calculated using Eq.~\ref{eq:singlet triplet yield} and~\ref{eq:singlet singlet yield}. Experiments involving the substitution of \ch{^{16}O} with \ch{^{17}O} are not new to researchers. Several such experiments have been performed in different biological contexts~\cite{paech2020quantitative,fiat1992vivo,mellon2009estimation}. Some of these experiments~\cite{mellon2009estimation,paech2020quantitative} have been performed with up to $70\%$ \ch{^{17}O} enrichment, making our assumption of one \ch{^{17}O} HFI reasonable.
\par
\begin{figure}[ht]
\centering
\includegraphics[scale=1]{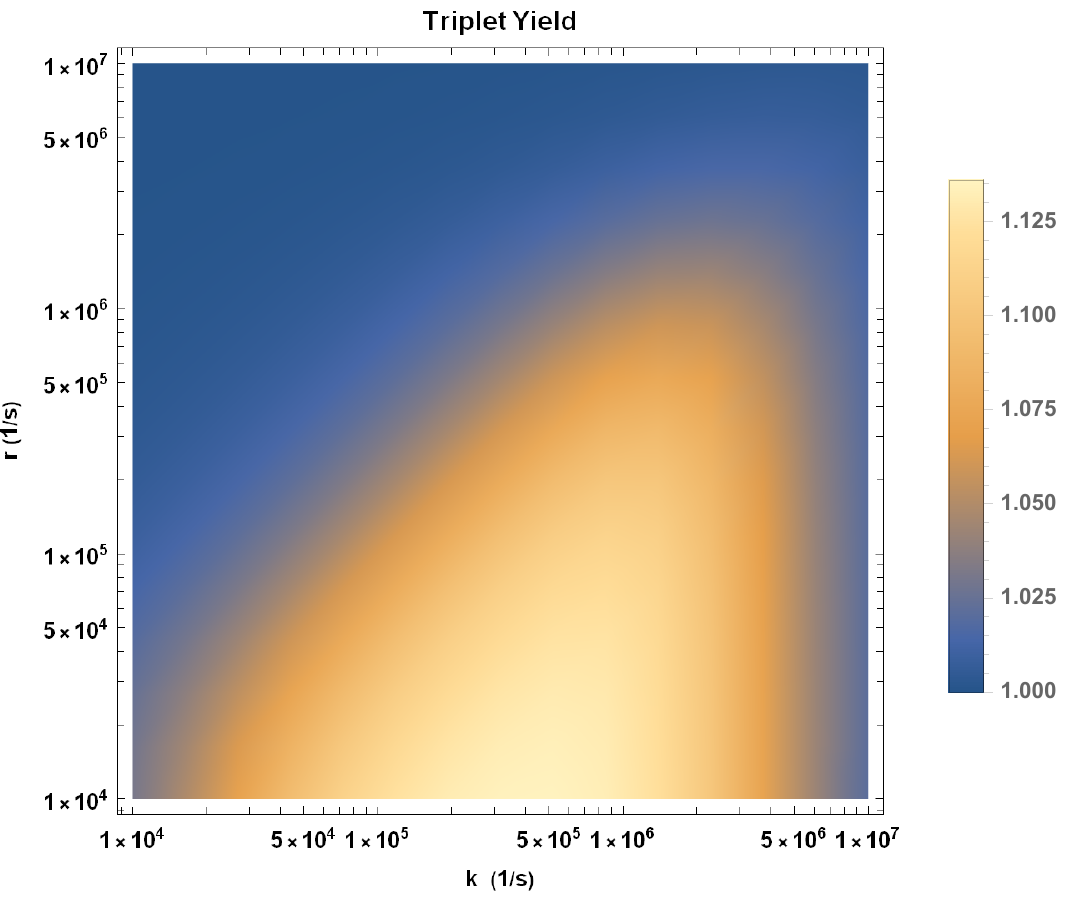}
\caption{{\bf Triplet yield ratio (GMF to HMF) for singlet-born RP containing \ch{^{17}O} in the $k-r$ plane.} The values of HFCCs are taken to be $a_{1} = -802.9$ $\mu T$~\cite{lee2014alternative} and $a_{2} = 1886.8$ $\mu T.$}
\label{fig:triplet O17}
\end{figure}

To understand the effects of this isotopic substitution, we plotted the triplet yield ratio (GMF to HMF) for RP containing \ch{^{17}O} on the $k$-$r$ plane (See Fig.~\ref{fig:triplet O17}). The comparison with Fig.~\ref{fig:triplet 1} reveals a significant difference: the strength of HMF effects is reduced on substituting \ch{^{16}O} with \ch{^{17}O}. A quantitative comparison for $k = 2\times10^{6}$ $s^{-1}$ and $r = 2\times10^5$ $s^{-1}$  is shown in Table~\ref{tab:O17 vs O16}.

\begin{table}[ht]
\centering
\caption{\label{tab:O17 vs O16} Percentage change in fractional triplet yield as one goes from GMF to HMF for singlet-born RP for both \ch{^{16}O} and \ch{^{17}O} cases. $a_{1} = -802.9$ $\mu T$, $a_{2} = 1886.8$ $\mu T$, $k = 2\times10^{6}$ $s^{-1}$, and $r = 2\times10^5$ $s^{-1}$.}
\begin{tabular}{|l|c|c|c|}
\hline
 Oxygen isotope & $\Phi_{T}$ at GMF & $\Phi_{T}$ at HMF & Percentage change \\
\hline
 \ch{^{16}O} & 0.596924 & 0.409039 & 31.4755
  \\
\hline
\ch{^{17}O} & 0.689092 & 0.630028 & 8.5714 \\
\hline
\end{tabular}
\end{table}

\section*{Discussion}
In this study, our principal goal was to answer whether the RPM could be the underlying mechanism behind the HMF effects on adult hippocampal neurogenesis as observed by Zhang et al.~\cite{zhang2021long}. Our results point to a possibility of such a mechanism. A simple RPM-based model with a set of typical parameters can reproduce the HMF effects of the same strength as those reported in the experimental findings of Zhang et al.~\cite{zhang2021long}. It is important to note that we have not modeled the neurogenesis process itself in this study, and that this would be an interesting direction to look at in future studies. 
\par
We proposed that the chemistry of \ch{[FADH^{.}... O2^{.-}]} RP may be responsible for the decrease in the concentration of superoxide in hypomagnetic conditions, which in turn results in adverse effects on adult hippocampal neurogenesis and related cognition. This RP and its role in ROS production have been implicated in the paper by Usselman et al.~\cite{usselman2016quantum}. A similar RP has been proposed in the past for avian magnetoreception~\cite{hogben2009possible,muller2011light,lee2014alternative}, the magnetic field effects on the circadian clock~\cite{zadeh2021radical}, and isotopic dependence of behavioral effects of Lithium~\cite{zadeh2021entangled}, where it is believed that lithium treatment modulates the oxidative stress level. We find that singlet-born RP qualitatively mimics the measurement of superoxide concentration by Zhang et al.~\cite{zhang2021long}.
\par
 Zhang et al.~\cite{zhang2021long} did not measure \ch{H2O2} concentration, any future measurement of \ch{H2O2} and \ch{O2^{.-}} under hypomagnetic conditions will be an essential check for our model.
\par
Zhang et al. show that the detrimental effects of long-term HMF exposure on adult neurogenesis were reversed by a pharmacological intervention to inhibit SOD1. Chemically, the dismutase activity of SOD1 accelerates the reaction of the superoxide anion  with itself to form hydrogen peroxide  and oxygen~\cite{wang2018superoxide}. Therefore, inhibition of SOD1 should increase the superoxide concentration in HMF exposed mice. This is in agreement with our assumption that the adverse effects of HMF on adult hippocampal neurogenesis are due to a decrease in the concentration of superoxide. Similarly, Zhang et al. also showed that inhibiting Nox-mediated ROS production when returning HMF-exposed mice back to GMF blocks the positive effect of a return to GMF on adult neurogenesis. Inhibition of Nox-mediated ROS production leads to a decrease in Superoxide production~\cite{terzi2020role} and therefore again agrees with our conclusion that adverse effects of HMF on adult hippocampal neurogenesis is due to a decrease in the concentration of superoxide.

\par
Results of our calculations favor singlet-born \ch{[FADH· · ·O2^{.-}]} RP rather than a triplet-born \ch{[FADH· · ·O2^{.-}]} RP. Mostly in RPs involving \ch{O2^{.-}} and \ch{FADH^{.}} the initial state of RP is taken to be a triplet state. It is assumed that in the RP formation process, the oxygen molecule prior to the electron transfer from flavin is in its triplet ground state, and consequently, the initial state of the the RP formed would be a triplet state. However, it is possible that the initial state of the oxygen molecule is the excited singlet state~\cite{saito1981formation,hayyan2016superoxide,kanofsky1989singlet,kerver1997situ,miyamoto2014singlet} (which is a biologically relevant ROS) instead. Moreover, the spin-orbit coupling could also transform the initial state of the RP from triplet to singlet states via intersystem crossing~\cite{goushi2012organic,fay2019radical}. Singlet initial state for RPs including superoxide as one of the component is consistent with assumptions of Refs~\cite{smith2021radical,zadeh2021entangled,zadeh2021radical}, and thus our results here strengthen their assumption of an initial singlet state.
\par
It has been suggested in the past that due to fast molecular rotation free \ch{O2^{.-}} has a spin relaxation lifetime on the orders of $1$ ns and hence a fast spin relaxation rate $r$~\cite{hogben2009possible,player2019viability}. The relaxation rate requirement calculated by our model yields $r$ significantly lower than this expected value. However, it has also been pointed out that this fast spin relaxation of free superoxide can be lowered if the molecular symmetry is reduced and the angular momentum is quenched by the biological environment~\cite{hogben2009possible,player2019viability}. It has also been proposed that fast spin relaxation of \ch{O2^{.-}} can be reduced by the involvement of scavenger species around \ch{O2^{.-}}~\cite{kattnig2017radical,kattnig2017sensitivity,babcock2021radical}.

\par
There is a possibility that other explanations may underlie the HMF effects on adult hippocampal neurogenesis. However, as our study shows, the outright rejection of RPM as by Zhang et al.~\cite{zhang2021long} does not seem justified, and RPM as a potential underlying mechanism should be taken seriously.
\par
Our RPM-based model for HMF effects on adult hippocampal neurogenesis implies the dependence of hippocampal neurogenesis on changes in the external magnetic field (See Fig.~\ref{fig:mag}). It would be interesting to conduct experiments to explore the impact of the static and oscillating external magnetic field on hippocampal neurogenesis. 
\par
Let us note that Pooam et al.~\cite{pooam2020hek293} reported that ROS production in HEK293 cells responds much more strongly to HMF than higher intensity static magnetic fields such as $500$ $\mu T$. As it is evident from Fig.~\ref{fig:mag}, the behavior of our model is in agreement with this observation, and we should expect effects of similar nature concerning magnetic field sensitivity of adult hippocampal neurogenesis.
\par
As singlet-triplet yield in an RPM can also be altered by isotopic substitution, it would be interesting to check our predictions regarding substitution of \ch{^{16}O} with \ch{^{17}O} in superoxide radical. We expect the strength of HMF effects to be reduced as a result of this substitution. Li is also known to affect hippocampal neurogenesis~\cite{fiorentini2010lithium}, it would also be interesting to observe any isotopic dependence for Li on hippocampal neurogenesis.

\par

It will be interesting to see whether the RPM-based explanation could be given for other known magnetic field effects in the brain, such as LI-rTMS, where it is known that magnetic fields and CRY play a central role~\cite{dufor2019neural}. It will also be interesting to see whether a role is played by the RPM in other contexts where ROS play a crucial role, such as mental disorders associated with oxidative stress~\cite{salim2014oxidative}. Van Huizen et al.~\cite{van2019weak} have reported that static weak magnetic  fields altered stem cell proliferation and differentiation via changes in reactive oxygen species (ROS) in planaria.  It will also be an interesting case for future studies to examine whether radical pairs play any role here.
\par
Our results also suggest that the quantum entanglement of the singlet state might be necessary for the mechanism of HMF effects on adult hippocampal neurogenesis. It has been suggested in the past that large-scale entanglement in the brain may underlie some neurobiological phenomena~\cite{hameroff2014quantum,hameroff2014consciousness,fisher2015quantum,simon2019can,adams2020quantum}. Entangled RP could be one of the sources of this entanglement, and our results are consistent with this idea in addition to Refs~\cite{smith2021radical,zadeh2021entangled}. For such a model to work, the exchange of quantum information between these RP is essential. It has been suggested that biophotons could serve as quantum messengers to establish long-distance connections~\cite{kumar2016possible,simon2019can}. In this context, it should also be noted that superoxide radicals can give rise to singlet oxygen, which is a potential source of biophotons~\cite{cifra2014ultra}.

\section*{Materials and methods}
\subsection*{RPM Calculations}
The state of the RP was described using the spin density operator. As shown in the results section (Eq.~\ref{eq:ham}), the spin hamiltonian involved only the Zeeman and the HF interactions. The exchange interaction of the electron spins and the dipolar interaction of the two electron spins were ignored~\cite{hore2016radical,efimova2008role,zadeh2021entangled,smith2021radical}.
\par
The Method of Timmel et al.~\cite{timmel1998effects,hore2019upper}  was followed to deal with the RP dynamics. The time dependence of the spin density operator in the absence of spin relaxation and chemical reactions was obtained from the von Neumann equation, and the chemical fate of the RP was modeled by means of separate first-order spin-selective reactions of the singlet and triplet pairs. Spin relaxation was introduced phenomenologically following Bagryansky et al.~\cite{bagryansky2007quantum}.
\par
The computational calculations and plotting were performed on Mathematica~\cite{Mathematica}.
\par
\subsection*{DFT Calculations}
We used the ORCA package~\cite{Neese2011} for the DFT calculation to obtain the HFCC of \ch{^{17}O} in superoxide. The molecular structure was optimized using PBE0/def2-TZVP. Using RI-B2GP-PLYP/def2-QZVPP~\cite{Goerigk2011}, we obtained $a_{\ch{^{17}O}}=1886.8$ $\mu$T. Relativistic effects were treated by a scalar relativistic Hamiltonian using the zeroth-order regular approximation (ZORA)~\cite{vanLenthe1996}. We considered the solvent effects using the conductor-like polarizable continuum model (CPCM)~\cite{Marenich2009}, with a dielectric constant of 2. 

\section*{Acknowledgments}
The authors would like to thank Jordan Smith, Wilten Nicola, and Douglas Wallace for their valuable input.



\end{document}